\documentclass[showpacs,preprintnumbers,aps]{revtex4}
\usepackage{epsf}
\usepackage{dcolumn}
\usepackage{bm}

\draft

\begin{document}

\title{Time-reversal violating generation of static magnetic and electric
fields and a problem of electric dipole moment measurement}
\author{ V. G. Baryshevsky}
\affiliation{Research Institute for Nuclear Problems, Belarusian State University, 11
Bobryiskaya str., 220050, Minsk, Republic of Belarus,\\
E-mail: bar@inp.minsk.by }
\date{\today}

\begin{abstract}
It is shown that in the experiments for search of EDM of an electron (atom,
molecule) the T-odd magnetic moment induced by an electric field and the
T-odd electric dipole moment induced by a magnetic field will be also
measured. It is discussed how to distinguish these contributions.
\end{abstract}

\maketitle


\narrowtext

Nowadays there is an appreciable progress the in development of methods for
ultraweak magnetic and electric field mesurement. {\ Therefore, new
experiments for measurement of an electric dipole moment $d$ (EDM) of
electrons (atoms, molecules) [1-4] are being prepared and carried out}.

The EDM of a particle exists if parity (P) and time-reversal (T) invariance
are violated. Investigation of the EDM existence could provide knowledge
about physics beyond the Standard Model [1-4].

F.L.Shapiro's idea \cite{5} to measure electron EDM by applying a strong
electric field to a subtance that has an unpaired electron spin is being
used for EDM search (see \cite{6,1}).

Interaction $W_E$ of an electric dipole moment $\vec{d}$ of an electron with
an electric field $\vec{E}$ depends on their orientation: 
\begin{equation}
W_E=-\vec{d}\vec{E},  \label{1}
\end{equation}
where $\vec{d}=d \frac{\vec{J}}{J}$, $\vec{J}$ is the atom spin, d is the
EDM.

Spins of electrons (atoms) at low temperature appear to be polarized similar
to the polarization (magnetization) of electrons by a magetic field in
paramagnetic subtances due to the interaction $W_B$ of an electron (atom)
magnetic moment $\vec{\mu}$ with a magnetic field $\vec{B}$ 
\begin{equation}
W_B=-\vec{\mu}\vec{B},  \label{2}
\end{equation}

Spins of electrons (atoms) polarized by an electric field induce the
magnetic field $\vec{B}_E$ and change in the magnetic flux $\Phi$ at the
surface of a flat sheet of material \cite{1}: 
\begin{equation}
\Delta \Phi=4 \pi \chi A d E^{*}/\mu_a,  \label{3}
\end{equation}
\begin{equation}
{B_E}=\frac{\Delta \Phi}{A}=4 \pi \chi \frac{d}{\mu_a}{E^{*}},  \label{4}
\end{equation}
where $\chi$ is the magnetic susceptibility, $\chi \approx \frac{\rho {\mu_a}%
^2}{3 k_{B} T}$, $\rho$ is the number density of spins of interest, $k_B$ is
Boltzmann's constant and $T$ is the sample temperature. In the cases where
simple Langevin paramagnetism is applicable, $E^{*}$ is the effective
electric field at the location of the spins, $\mu_{a}=g \sqrt{J(J+1)}\mu_{B}$
where $\mu_{B}$ is the Bohr magneton, $\mu_{a}$ is the atomic or ionic
magnetic moment, $g$ is Lande factor and $A$ is the sample area.

If an external magnetic field acts on either a para- or a ferromagnetic
material, the spins in a substance become polarized due to substance
magnetization. Therefore, the electric dipole moments appears polarized too.
This results in the induction of an electric field $\vec{E}_B$ 
(D.DeMille [1]): 
\begin{equation}
{E_B}=4 \pi \rho d P(B),  \label{5}
\end{equation}
where $P$ represents the degree that the spins are polarized in the sample.

According to the analysis \cite{1}, modern methods for measurement of ${B_E}$
and ${E_B}$ provide sensitivity for electric dipole moment measurement about
10$^{-32}~e~cm$ and in some cases even 10$^{-35}~e~cm$.

It is important to pay attention to another mechanism of time-reversal
violating generation of magnetic and electric fields, which have been
discussed in \cite{7,8}. According to the idea of \cite{7,8}, an induced
magnetic moment ${\vec \mu}({\vec E})$ of a particle appears due to the
action of a field $\vec E$ under conditions of violation of P- and
T-invariance (and similar, an induced electric dipole moment ${\vec d}_B$ of
a particle appears due to the action of a field $\vec B$). This new effect
does not depend on temperature. An effect magnitude is determined by a P-odd
T-odd tensor polarisability $\beta_{ik}^{T}$ of a particle (atom, molecule,
nucleus, neutron, electron and so on). For an atom (molecule), $%
\beta_{ik}^{T}$ arises due to P- and T-odd interaction of electrons with a
nucleus. %
%
For the stationary state of an atom (molecule) $|N_{0}\rangle$ the tensor $%
\beta_{ik}^{T}$ is as follows: 
\begin{equation}
\beta _{ik}^{T}=\sum_{F}\frac{\langle N_{0}|\widehat{d}_{i}|F\rangle
\;\langle F|\widehat {\mu} _{k}|N_{0}\rangle +\langle N_{0}|\widehat {\mu}%
_{i}|F\rangle \;\langle F|\widehat {d}_{k}|N_{0}\rangle }{E_{F}-E_{N_{0}}},
\end{equation}
where $|F\rangle$ is the wave function of a stationary state of the atom,
considering T-odd interaction $V_w^T$, $E_F$ and $E_{N_0}$ are the energies
of the atom (molecule) stationary states, $\widehat{\overrightarrow{d}}$ and 
$\widehat{\overrightarrow{\mu}}$ are the operators of electric dipole moment
and magnetic moment, respectively and $i,k=1,2,3$ correspond to the axes $%
x,y,z$.

Let us place an atom (molecule) into an electric field $\vec {E}$. The
induced magnetic dipole moment $\vec{\mu} (\vec E)$ appears in this case %
\cite{7,8} : 
\begin{equation}
{\mu_{i} (\vec E)}=\beta_{ik}^{T} {E}_k,
\end{equation}

The tensor $\beta_{ik}^{T}$ (like any tensor of rank two) can be expanded
into scalar, simmetric and antisimmetric parts.

The antisymmetric part of the tensor $\beta_{ik}^{T}$ is proportional to $%
e_{ikl} J_l$, where $e_{ikl}$ is the totally antisymmetric tensor of rank
three. The symmetric part of the tensor $\beta_{ik}^{T}$ is proportional to
the tensor of quadrupolarization $Q_{ik}=\frac{3}{2J(2J-1)}[J_i J_l+J_k J_l-%
\frac{2}{3} J(J+1)\delta_{ik}]$. As a result 
\begin{equation}
\beta_{ik}^{T}=\beta_{s}^{T} \delta_{ik} + \beta_{v}^{T} e_{ikl} J_l +
\beta_{t}^{T} Q_{ik},
\end{equation}
where $\beta_{s}^{T},\beta_{v}^{T}, \beta_{t}^{T}$ are the scalar, vector
and tensor P-, T-odd polarizabilities of the particle, respectively. For a
substance with the nonpolarized spins $Sp~\rho(J){\vec J}=0$ and $Sp~\rho
(J) Q_{ik}=0$ (here $\rho (J)$ is the atom (molecule) spin density matrix).
As a result for such a substance, $\beta_{ik}^{T}$ appears to be a scalar $%
\beta_{ik}^{T}=\delta_{ik} \beta_{s}^{T}$.

Placement of a nonpolarized atom (molecule, nucleus) 
into an electric field induces the magnetic dipole moment $\vec{\mu}_E$ (see
also Appedix): 
\begin{equation}
\vec{\mu} (\vec E)=\beta_{s}^{T} \vec{E},  \label{8}
\end{equation}
where $\beta _{s}^{T}=\sum_{F}\frac{\langle N_{0}|\widehat{d}_{z}|F\rangle
\;\langle F|\widehat{\mu} _{z}|N_{0}\rangle +\langle N_{0}|\widehat{\mu}
_{z}|F\rangle \;\langle F|\widehat{d}_{z}|N_{0}\rangle }{E_{F}-E_{N_{0}}}$, $%
\widehat{d}_{z}$ and $\widehat{\mu}_{z}$ are the $z$ components of the
operators of the electric dipole moment and magnetic moment, respectively
with axis $z$ parallel to the electric field $\vec{E}$. It should be
emphasized that for strong fields (when the distance between atom (molecule)
levels is comparable with the energy of interaction with an electric $\vec E$
(magnetic $\vec B$) field) $\beta_{s}^{T}$ depends on ${\vec E}~({\vec B})$. 
Weak interaction is much weaker than strong and elecromagnetic interactions.
Therefore, to find the wave function $|F\rangle$, the perturbation theory
can be applied: 
\begin{equation}
|F\rangle=|f\rangle+\sum_{n} \frac{\langle n|V_{w}^T|f\rangle} {E_{f}-E_{n}}%
|n\rangle= |f\rangle+\sum_{n} \eta_{nf}|n\rangle ,  \label{7}
\end{equation}
where $|f\rangle$ is the wave function of an atom in the absence of weak
interactions and the mixing ratio is $\eta_{nf}=\frac{\langle n|W|f\rangle}{%
E_{f}-E_{n}}$. It should be mentioned that for theoretical analysis of $%
\beta_{s}^T$ in a substance it is neccessary to find a wave function of an
excited state of an atom in the substance which is difficult to do. 

It follows from (\ref{8}) that in a substance placed into electric field the
magnetic field is induced \cite{7,8}: 
\begin{equation}
\vec{B}_{E}^{ind}=4 \pi \rho \beta _{s}^{T} \vec{E}^{*}.  \label{9}
\end{equation}

Vice versa, if an atom (molecule, nucleus) is placed into a magnetic field,
the induced electric dipole moment $\vec{d}_B$ appears \cite{7,8}. 
The tensor polarizability associating $\vec d$ with $\vec B$ is $%
\chi_{ik}^T=\beta_{ki}^T$. If only contribution from the scalar
polarizability is significant, then induced EDM can be represent as follows: 
\begin{equation}
\vec{d}_B=\beta_{s}^{T} \vec{B}^{*},  \label{8-2}
\end{equation}
which leads to the induction of an electric field in the substance: 
\begin{equation}
\vec{E}_{B}^{ind}=4 \pi \rho \beta _{s}^{T} \vec{B}^{*} ,  \label{10}
\end{equation}
where $B^{*}$ ($E^{*}$) is the local field, acting on the considered
particle in the substance.

Hence, analysing the results of the experiment proposed in \cite{1}, one
should consider that the appearance of the induced magnetic and electric
fields is caused by: 

\begin{enumerate}
\item A magnetic field is induced due to interaction of the electric dipole
moment of an atom with an external electric field \cite{1,5} (see (\ref{3}),(%
\ref{4}))

\item A magnetic field is induced due to mechanism \cite{7,8} (see (\ref{9}%
)) 

\item A magnetic field appears as a result of polarization (magnetization)
of atom magnetic moments by the local induced magnetic field ${\vec B}_{loc}$
due to interaction $W$ of the magnetic dipole moment of an atom with this
field 
\begin{equation}
W=-{\vec{\mu}_a} {\vec B}_{loc}.
\end{equation}
The local field ${\vec B}_{loc}$ is the sum of two contributions: 
\begin{equation}
{\vec B}_{loc}={\vec B}_{E~loc}+{\vec B}_{loc}^{ind},
\end{equation}
where the field ${\vec B}_{E~loc}$ is the local magnetic field acting on an
atom from the polarized (by mechanism [1,5], see (3),(4)) magnetic moments
of the other atoms of the sample. This field depends on temperature and its
contribution could be neglected for those temperature values, which provide $%
\chi \ll 1$. But for temperature $T<1K$ the susceptibility $\chi \sim \frac{1%
}{T}$ becomes comparable with $1$ and higher, and the energy of interaction
of two magnetic dipoles for neighbour atoms occurs of order of $k_{B} T$ and
greater. Thus, in this case, the collective effects, well-known in the
theory of phase transition in magnetism, should be taken into account while
considering magnetization by an electric field.

The field ${\vec B}_{loc}^{ind}={\vec B}_{1~loc}^{ind}+{\vec B}%
_{2~loc}^{ind} $ does not depend on temperature.

The field ${\vec B}_{1~loc}^{ind}=\chi_{1~loc}^T {\vec E}^{*}$ is the local
magnetic field produced in the point of the considered atom location by the
magnetic moments of atoms of the substance (except for the considered atom)
induced by the aid of mechanism [7,8] (see (6),(8)); $\chi_{1~loc}^T \sim
\rho \beta _{s}^{T}$ is the local P-,T-odd susceptibility of the substance
(it depends on the substance density and sample shape: for sphere $%
\chi_{1~loc}^T=\frac{8 \pi}{3} \rho \beta _{s}^{T}$ for cylinder $%
\chi_{1~loc}^T={4 \pi} \rho \beta _{s}^{T}$ ).

The field ${\vec B}_{2~loc}^{ind}$ is the self-induced magnetic field of the
considered atom. The magnetic moment (T-odd current) of the atom induced by
an electric field acting on the atom due to mechanism [7,8] causes
appearance of the magnetic field inside the atom: 
\begin{equation}
{\vec H}_{E}^{T}({\vec r})=rot~{\vec A}_{E}^{T}({\vec r}),
\end{equation}
with vector potential 
\begin{equation}
{\vec A}_{E}^{T}({\vec r})=\frac{1}{c} \int \frac {j_{E}^{T}({\vec r}%
^{\prime})} {\left|{\vec r}-{\vec r}^{\prime})\right|} d^3 r^{\prime},
\end{equation}
$j_{E}^{T}({\vec r}^{\prime})$ is the T-odd part of operator of the current
density for an atom (molecule) placed in an electric field (it is calculated
by the use of wavefunctions similar (10) (\cite{12,new_lanl})). The magnetic
interaction hamiltonian of an atom with the field ${\vec A}_{E}^{T}$ can be
expressed as follows (see Appendix 2): 
\begin{equation}
W_{2~loc}=-\frac{1}{2c} \int ( {\vec j}^0({\vec r}) {\vec A}^{T}({\vec r})+ {%
\vec A}^{T}({\vec r}) {\vec j}^0({\vec r}) ) d^3 r,
\end{equation}
where ${\vec j}^0({\vec r})$ is the atom current density operator calculated
with the atom wavefunction without consideration of P-,T-odd interactions, 
\begin{equation}
{\vec j}^0({\vec r})=c~rot{\vec {\mu}}({\vec r}),
\end{equation}
and ${\vec {\mu}}({\vec r})$ is the operator of magnetic moment density.

The above expressions allow us to rewrite $W_2$  as follows: 
\begin{eqnarray}
W_{2~loc}= -{\vec {\mu}}_a {\vec B}_{loc}^{ind}=-{\mu}_{ai}{\chi_{(at)}^T}%
_{ik} E_{k}^{*},  \nonumber
\end{eqnarray}
where ${\vec {\mu}}_a$ is the operator of the magnetic moment of an atom, ${%
\vec {\mu}}_a={\mu}_a \frac{{\vec J}}{J}$ and $\chi_{(at)ik}^T$ is the
tensor of T-odd atom susceptibility, which does not depend on the substance
density and sample shape. The scalar part $\chi_{at}^T$ of the T-odd atom
susceptibility tensor is $\chi_{at}^T \sim \beta _{s}^{T} \frac{1}{a^3}$
(here $a$ is the typical radius of distribution density of the magnetic
moment induced by an electric field in the atom \cite{7,8}). 

As a result 
\begin{equation}
{\vec B}_{loc}^{ind}=(\chi_{1~loc}^T + \chi_{at}^T) {\vec E}%
_{loc}^{*}=\chi_{loc(subst)}^T {\vec E}_{loc}^{*},
\end{equation}
\begin{equation}
\chi_{loc(subst)}^T=\chi_{1~loc}^T + \chi_{at}^T.
\end{equation}
%
%
\end{enumerate}

%
%
The interaction (11) of the magnetic moment of an atom with the induced
magnetic field causes the appearance of the magnetic field due to different
population of magnetic levels of the atom in the field ${\vec B}_{loc}$ in
thermal equilibrium 
\begin{equation}
{\vec B}^{\prime~ind}= 4 \pi \chi {\vec B}_{loc} \approx 4 \pi \chi {\vec B}%
_{loc}^{ind}= 4 \pi \frac{\rho \mu_{a}^{2}}{3 k_{B} T} \chi_{loc(subst)}^T {%
\vec E}^{*},
\end{equation}
%
the field ${\vec B}_{loc}={\vec B}_{E~loc}+{\vec B}_{loc}^{ind}$, but ${\vec
B}_{E~loc}$ contribution could be neglected for those temperature values,
which provide $\chi \ll 1$, and it is omitted here. 

Therefore, the flux $\Delta \Phi$, which is going to be measured in the
experiment proposed in \cite{1} should be written as: 
\begin{eqnarray}
\Delta \Phi & = & A B_{E}= 4 \pi A (\chi \frac{d}{\mu_a}+ \rho \beta
_{s}^{T}+ \chi \chi_{loc(subst)}^T)E^{*} =  \nonumber \\
& = & 4 \pi A [\chi (\frac{d}{\mu_a}+\chi_{loc(subst)}^T) +\rho \beta
_{s}^{T}]E^{*} ,  \label{11}
\end{eqnarray}
\begin{eqnarray}
{\vec B}_E & = & 4 \pi (\chi \frac{d}{\mu_a}+ \rho \beta _{s}^{T}+ \chi
\chi_{loc(subst)}^T) {\vec E}^{*} =  \nonumber \\
& = & 4 \pi [\chi (\frac{d}{\mu_a}+\chi_{loc(subst)}^T) +\rho \beta
_{s}^{T}]E^{*},  \label{12}
\end{eqnarray}
where $\chi=\frac{\rho \mu_{a}^2}{3 k_{B}T}$.

The electric field measured in the experiment \cite{1} is as follows (see
(9),(10)): 
\begin{equation}
{E_B}=4 \pi \rho (d P(B)+ \beta _{s}^{T} B^{*}).  \label{13}
\end{equation}
The term proportional to the interaction of the electric dipole moment of an
atom with the electric field induced by mechanism \cite{7,8} is small (it is
of the second order over T-odd interaction) and is neglected.

Thus, measurement of $\Delta \Phi$ and ${E_B}$ provides knowledge about the
sum of quantities $d$, $\chi_{loc(subst)}^T$ and $\beta _{s}^{T}$. To
distinguish these contributions one should consider the fact that $\chi$ and 
$P(B)$ depend on temperature, while $\beta _{s}^{T}$ does not. Therefore,
studying $B_E$ and $E_B$ dependence on temperature allows one to evaluate
contributions from $d$, $\chi_{loc(subst)}^T$ and $\beta _{s}^{T}$ to the
measured effect.

It should be particularly emphasized that $\Delta \Phi$ and ${E_B}$ differs
from zero even when EDM $d$ is equal to zero.

According to \cite{1} we can expect a magnetic induction sensitivity about $%
3 \times 10^{-15}~G/ \sqrt{Hz}$. In ten days of averaging the sensitivity is 
$\sim 10^{-18}~G$. This leads to the sensitivity of about $10^{-32}~e~cm$
for $d$. 
Such sensitivity of magnetic induction measurement provides for
polarizability measurement the sensitivity $\beta _{s}^{T} \sim
10^{-43}~cm^{3}$ (see (\ref{12})) and for mixing ratio the sensitivity $%
\eta_T \sim 10^{-17}$. 
This value for $\eta_T$ is significantly lower than the limitation for
mixing ratio $\eta_T \sim 10^{-14}$ [7-10], which could be obtained from the
results of measurements of atom dipole moment that have been done earlier
(see, for example, \cite{9}).

It should be emphasized that the polarizability $\beta _{s}^{T}$ and the
susceptibility $\chi_{loc(subst)}^T$ differs from zero even for atoms with
the zero spin, for which EDM is absent.

\section*{Conclusion}

In \cite{1} the experiments for search of the EDM of an electron (atom) by
measurement of induced magnetic and electric fields are discussed. In the
present paper it is shown that in this experiments the T-odd magnetic moment
induced by an electric field and the T-odd electric dipole moment induced by
a magnetic field will be also measured. The T-odd scalar polarizability $%
\beta _{s}^{T}$ and the susceptibility $\chi_{loc(subst)}^T ~(\chi_{at}^T)$
will be measured too. It is shown that study of temperature dependence of $%
\Delta \Phi$ and ${E_B}$ allows us to distinguish the contribution provided
by $\chi_{loc(subst)}^T ~(\chi_{at}^T)$ and the electric dipole moment $d$
provided by $\beta _{s}^{T}$.

Since external electric and magnetic fields are used in the experiments
[2-4], then the described mechanism of induction of magnetic and electric
fields \cite{8} should be taken into account for understanding the EDM
contribution in the above experiments.

For example, in \cite{2} it is proposed to detect EDM by measurement of
Zeeman precession frequency of $^{199}$Hg nuclear spins in parallel electric 
${\vec E}$ and magnetic ${\vec B}$ fields. The measurements are
simultaneously performed in two cells with oppositely directed electric
fields to reduce the frequency noise due to magnetic field fluctuations.
According to \cite{2} a difference between the Zeeman frequencies in these
two cells is 
\begin{equation}
\hbar (\omega_1 - \omega_2)=4 d E.  \label{c1}
\end{equation}
But it should be considered that an electric field induces magnetic moment
and, consequently, the additional local magnetic field 
\[
{\vec B}_{loc}^{ind}=\chi_{loc(subst)}^T \vec{E}. 
\]
acts on the nucleus spin.

The magnetic moment of a Hg atom interacts with this field: 
\begin{equation}
W=-{\vec \mu}_{Hg}{\vec B}_{loc} =-{\vec \mu}_{Hg} \chi_{loc(subst)}^T \vec{E%
}.  \label{c2}
\end{equation}
$\chi_{loc(subst)}^T$ is the local P-,T-odd susceptibility of the substance.

Vice versa, the magnetic field $\vec{B}$ induces the electric dipole moment,
which creates the electric field that interacts with the EDM (if it exists).
But this effect contributes only a little bit to the frequency difference,
because it is of the second order over T-odd interaction.

Thus, frequency difference can be presented as follows: 
\begin{equation}
\hbar (\omega_1 - \omega_2)=4 (d + \mu_{Hg} \chi_{loc(subst)}^T) E.
\label{c3}
\end{equation}

Note that for gases $\chi_{loc(subst)}^T \approx \chi_{at}^T$.

It is evident that the experimentally observed value again includes
contributions from two effects: EDM and the effect described in \cite{7,8}.

The similar situation is for \cite{3}. Here the parameter $\chi_{at}^T$
appears as an additional term in the coefficient $\varepsilon$ (see eq.(2)
in \cite{3}).

For the experiment with molecules \cite{4} an external electric field also
induces the magnetic moment (by the mechanism described in \cite{7,8}), due
to which the field $\vec{B}+{\mu}_{B} \chi_{at}^T {\vec E}$ acts on the
magnetic moment of the molecule instead of the external field $\vec{B}$.

Therefore, the difference in phase shifts $2 \phi$ considered in \cite{4}: 
\[
2 \phi=(2 \mu_{B} B + d_e \eta E )\frac{T}{\hbar} 
\]
should be replaced by 
\begin{equation}
2 \phi=[2 \mu_{B} B + (d_e \eta + 2 \mu_{B} \chi_{at}^T) E ]\frac{T}{\hbar}.
\label{c4}
\end{equation}
Note that contribution from $\chi_{at}^{T}$ in $2 \phi$ differs from zero
even if an electron's EDM $d_e$ is equal to zero.

It is also important to draw attention that in the interferometric method %
\cite{4} molecules pass through external electric and magnetic fields. 
{Hence, this experimental setup practically realizes the experimental setup
proposed in \cite{13} for the method of atomic spin interferometry.}
According to \cite{13}, even interactions, which do not depend on spin
orientation, {contribute to the spin rotation in a field {\ providing the
phase difference $2 \phi$ between different spin states.} 
For example, conventional electromagnetic interaction $U=-\frac{1}{2} \alpha
E^2$ also contributes in the difference of phase shifts $2 \phi$ (here $%
\alpha$ is the conventional P-,T-invariant polarizability, which does not
even depend on spin orientation). It means that one should change the sign
of the electric field to eliminate contribution proportional to $E^2$ and
other similar contributions (scattering by the residual gas in setup and
even interaction of a molecule with gravitational field). }

It should be noted that the magnetic field induced by an electric field
appears on electron due to its possible polarizability. From the
dimensionality consideration it follows that this contribution to the levels
splitting $W_e=-{\overrightarrow{\mu}}_e \chi_e \overrightarrow{E}$ is
comparable with $W=-{\overrightarrow{d}}_e \overrightarrow{E}$.

And similar, the magnetic field induced by an electric field
appears on nucleus due to its possible polarizability. From the
dimensionality consideration it follows that this contribution to the levels
splitting $W_{nucl}=-{\overrightarrow{\mu}}_{nucl} \chi_{nucl} \overrightarrow{E}$ is
comparable with $W=-{\overrightarrow{d}}_{nucl} \overrightarrow{E}$.

Attention also should be paid to the fact that elementary particles also can
possess nonzero susceptibility $\chi $. As an example, let us consider an
experiment for measurement of electric dipole moment of neutron. Suppose
neutron is placed into an electric field. If neutron possesses electric
dipole moment, then, at first sight, the energy of interaction of this
dipole moment with the electric field is $W_{n1}=-d_{n}\overrightarrow{%
\sigma }\overrightarrow{E}$, where $d_{n}$ is the electric dipole moment of
the neutron, $\overrightarrow{\sigma }$ are the Pauli matrices. But, in
accordance with the above, the electric field induces the magnetic field on
the neutron $\overrightarrow{B_{ind}}=\chi _{n}\overrightarrow{E}$ and the
magnetic moment of the neutron interacts with this magnetic field $W_{n2}=-{%
\overrightarrow{\mu }}_{n}\chi _{n}\overrightarrow{E}=-{\mu }_{n}\chi _{n}%
\overrightarrow{\sigma }\overrightarrow{E}$ As a result, total energy of
interaction of a neutron with an electric field is: 
\begin{equation}
W_{n}=-(d_{n}+{\mu }_{n}\chi _{n})\overrightarrow{\sigma }\overrightarrow{E}
\label{Wn}
\end{equation}%
It is evident that even at $d_{n}=0$ $W_{n}$ differs from zero. It means
that in experiments for dipole electric moment measurement the sum $(d_{n}+{%
\mu }_{n}\chi _{n})$ will be measured. Experimental estimations for electric
dipole moment give value $d_{n}\sim 10^{-26}\div 10^{-27}$. But according to
(\ref{Wn}) this value concerns the sum $(d_{n}+{\mu }_{n}\chi _{n})$ i.e. $%
(d_{n}+{\mu }_{n}\chi _{n})\leq 10^{-26}\div 10^{-27}$. Then, considering $%
d_{n}=0$, the limit for $\chi _{n}$ can be obtained: $\chi _{n}=\frac{%
(10^{-26}\div 10^{-27})e}{\mu _{n}}\leq 10^{-13}$.

\section*{Appendix 1}

To illustrate the mechanism of the magnetic field induction by an electric
field let us consider a simple model at first \cite{12}. Suppose an atom is
in the $s_{1/2}$ state and we place it to an electric field. Taking into
account the admixture of the nearest $p_{1/2}$ state due to $P$- and $T$-odd
interactions of an electron with a nucleus and interaction with the electric
field one can represent the wave function of an atom in the form: 
\begin{eqnarray}
|{\ \widetilde s_{1/2} } \rangle &=& \frac{1}{\sqrt{4 \pi}} [R_0 (r) - R_1
(r) ({\vec \sigma} \vec n) \eta_T \\
& & - R_1(r) ({\vec \sigma} {\vec n}) ({\vec \sigma} {\vec E}) \delta ]
|\chi_{1/2} \rangle  \nonumber
\end{eqnarray}
Here $\vec \sigma$ are the Pauli matrices, $\vec n = \vec r /r$ is the unit
vector along $\vec r$, $\vec E$ is the electric field strength, $R_0$ and $%
R_1$ are radial parts of $s_{1/2}$ and $p_{1/2}$ wave functions
respectively, $| \chi_{1/2} \rangle$ is the spin part of wave function, 
$\eta_T$ is the mixing coefficient describing $P$ and $T$ noninvariant
mixing of S and P states, $\delta$ describes mixing of S and P states by the
Stark effect and $\vec E$ is the electric field strength.

Interference of Stark and $PT$ - odd terms changes the electron spin
direction $\Delta \vec s$ in the point $\vec r$ as follows: 
\begin{eqnarray}
\Delta \vec s (\vec r) & = & \frac{ \eta_T \delta }{8 \pi} R_1^2 (r) \langle
\chi_{1/2} | (\vec \sigma \vec n) \vec \sigma (\vec \sigma \vec n) (\vec
\sigma \vec E)  \nonumber \\
& & + (\vec \sigma \vec E) (\vec \sigma \vec n) \vec \sigma (\vec \sigma
\vec n) | \chi_{1/2} \rangle  \nonumber \\
& = & \frac{\eta_T \delta R_1^2(r)}{ 8 \pi} \biggl( 4 \vec n (\vec n \vec E)
- 2 \vec E \biggr)
\end{eqnarray}
The vector field $4 \vec n (\vec n \vec E) - 2 \vec E $ is shown in Fig. 1.
Since $\Delta \vec s$ does not depend on the initial direction of the atomic
spin, this spin structure appears even in a nonpolarized atom. The spin
vector averaged over spatial variables differs from zero and is directed
along $\vec E$. As a result, the magnetic moment of an atom also differs
from zero.

\begin{figure}[h]
\epsfxsize =7cm \centerline{\epsfbox{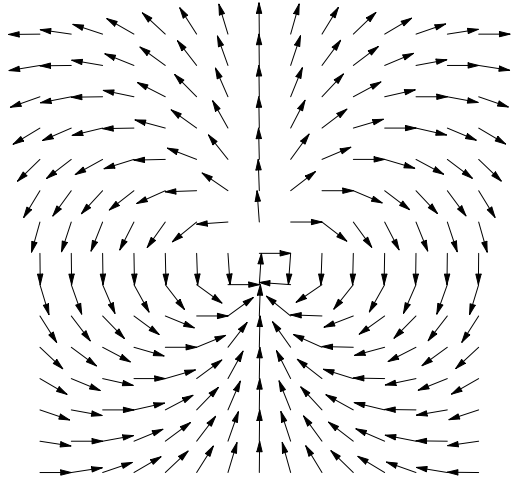}} 
\caption{\textit{Vector field $4 \vec n (\vec n \vec E) - 2 \vec E$. Vectors
on figure show direction of atomic spin in $s_{1/2}$ state if we take into
account the admixture of the $p_{1/2}$ state due to $PT$ noninvariant
interactions and external electric field.}}
\label{figb}
\end{figure}

\section*{Appendix 2}

Considering matrix elements $W$ for different orientations of spin of an
atom (molecule) with respect to the direction of electric field one can
obtain the shift of atom (molecule) levels depending on spin orientation.

The above semiclassical reasoning provides to obtain an expression for the
shift of atom (molecule)levels, which can be defind more accurately and more
exactly from analysis of quantum-electrodynamic radiation correction to the
levels of an atom (molecule) placed into an external field.

In this case the hamiltonian of the system contains not only the hamiltonian
of the atom in the external field $H^T_A$, but also the hamiltonian of
photons $H_{\gamma}$ and hamiltonianof interaction of the atom with photons $%
H_{A\gamma}$ 
\begin{eqnarray}
H=H^T_A+H_{\gamma}+H_{A\gamma}, \\
H^T_A=H_{0A}+V_{ext}+V_w =H^{ext}_A+V_w ,
\end{eqnarray}
where $H_{0A}$ is the hamiltonian of an atom (molecule) in the absence of
external fields and weak interactions, $V_{ext}$ is the energy of
interaction of an atom (molecule) with an external field, $V_w$ is the
energy weak interaction of electrons with nucleus and with each other, $%
H^{ext}_A=H_{0A}+V_{ext}$.

The studied induced levels shift in external fields appears 
in the second order of 
quantum electrodynamic perturbation theory over interaction of atoms with
photons $H_{A \gamma}$.

Following the receipt \cite{15} one can obtain that the magnitude of
radiation shift $\delta \varepsilon _{N}$ of a level is defined by an
expression similar to formula obtained in \cite{15} even in presence of
T-odd interactions: 
\begin{equation}
\delta \varepsilon _{N}=\frac{e^{2}}{16\pi ^{3}}P{\int }\frac{d^{3}k}{\left| 
\overrightarrow{k}\right| }\sum\limits_{F}\frac{\overline{\Psi }_{N}\left( 
\overrightarrow{r}\right) \gamma _{\mu }e^{i\overrightarrow{k}%
\overrightarrow{r}}\Psi _{F}\left( \overrightarrow{r}\right) \overline{\Psi }%
_{F}\left( \overrightarrow{r}^{\prime }\right) \gamma _{\mu }e^{-i%
\overrightarrow{k}\overrightarrow{r}}\Psi _{N}\left( \overrightarrow{%
r^{\prime }}\right) }{E_{N}-E_{F}-\frac{E_{F}}{\left| E_{F}\right| }\left| 
\overrightarrow{k}\right| }d^{3}rd^{3}r^{\prime },  \label{Pint}
\end{equation}%
where $\Psi _{N}$ are the eigenfunctions of Hamiltonian $H_{A}^{T}$ 
\begin{eqnarray}
H_{A}^{T}\Psi _{N} &=&E_{N}\Psi _{N},  \nonumber \\
\overline{\Psi }_{N} &=&{\Psi }_{N}^{+}\beta ;~{\gamma _{\mu },\beta ~\text{%
are the Dirac matrices, }\mu =1,2,3,4.}  \nonumber
\end{eqnarray}

The symbol $P{\int}$ denotes the principal value of integral.

Let us consider the shift of the level $N$ of the ground state of an atom
(molecule). Integrating (\ref{Pint}) over $d^3 k$ one obtains 
\begin{eqnarray}
\delta \varepsilon _{N}=\frac{1}{8\pi }\sum\limits_{N^{\prime }}\int \frac{%
j_{NN^{\prime }\mu }\left( \overrightarrow{r}\right) j_{N^{\prime }N\,\mu
}\left( \overrightarrow{r}^{\prime }\right) }{\left| \overrightarrow{r}-%
\overrightarrow{r}^{\prime }\right| }d^{3}rd^{3}r^{\prime } + \\
+ \frac{1}{4 {\pi}^2 }\sum\limits_{F\neq N}\int j_{NF\mu }\left( \overrightarrow{r%
}\right) \Phi _{FN}(\left| \overrightarrow{r}-\overrightarrow{r}^{\prime
}\right| )j_{FN\,\mu }\left( \overrightarrow{r}^{\prime }\right)
d^{3}rd^{3}r^{\prime },  \nonumber  \label{deN}
\end{eqnarray}
here $\sum\limits_{N^{\prime }}$ is the sum over degenerate spin states of
the level $N$; 

\noindent transition current 
\begin{eqnarray}
j_{NF{\mu}}(\overrightarrow{r})=i~e \overline{\Psi}_{N}\left(\overrightarrow{%
r}\right) \gamma _{\mu} {\Psi}_{N} \left(\overrightarrow{r}\right), \\
{\Phi}_{FN}(\left| \overrightarrow{r}-\overrightarrow{r}^{\prime }\right|)=
- \frac{4 \pi}{(\left| \overrightarrow{r}-\overrightarrow{r}^{\prime
}\right|)} \int\limits_{0}^{\infty }\frac{\sin k\left| \overrightarrow{r}-%
\overrightarrow{r}^{\prime }\right| }{E_{F}-E_{N}+k}dk, {\text{ for} E_F > 0}%
,  \label{-} \\
\Phi_{FN}(\left| \overrightarrow{r}-\overrightarrow{r}^{\prime }\right|)=  
\frac{4 \pi}{(\left| \overrightarrow{r}-\overrightarrow{r}^{\prime }\right|)}
\int\limits_{0}^{\infty }\frac{\sin k\left| \overrightarrow{r}-%
\overrightarrow{r}^{\prime }\right| }{E_{F}+E_{N}+k}dk, {\text{for} E_F < 0}
\label{+}
\end{eqnarray}

Integrals in (\ref{-}), (\ref{+}) can be computed explicitly 
\begin{equation}
\int\limits_{0}^{\infty } \frac{\sin \alpha x}{x+\beta}=ci(\alpha\beta)sin
(\alpha\beta)- cos(\alpha\beta)si(\alpha\beta), ~~~~ \left| arg~\beta \right|<
\pi,~ \alpha>0.
\end{equation}

Let us consider the fist term in (\ref{deN}) more attentively 
\begin{eqnarray}
{\delta }{\varepsilon }_{N}^{(1)} &=&\frac{1}{8\pi }\int \frac{j_{NN^{\prime
}\mu }\left( \overrightarrow{r}\right) j_{N^{\prime }N\,\mu }\left( 
\overrightarrow{r}^{\prime }\right) }{\left| \overrightarrow{r}-%
\overrightarrow{r}^{\prime }\right| }d^{3}rd^{3}r^{\prime }= \\
&=&\frac{1}{8\pi }\int \frac{\overrightarrow{j}_{NN^{\prime }}(%
\overrightarrow{r})\overrightarrow{j}_{N^{\prime }N}(\overrightarrow{%
r^{\prime }})}{\left| \overrightarrow{r}-\overrightarrow{r}^{\prime }\right| 
}d^{3}rd^{3}r^{\prime }-  \nonumber \\
&-&\frac{1}{8\pi }\int \frac{\Psi _{N}^{+}\left( \overrightarrow{r}\right)
\Psi _{N^{\prime }}\left( \overrightarrow{r}\right) \Psi _{N^{\prime
}}^{+}\left( \overrightarrow{r}^{\prime }\right) \Psi _{N}\left( 
\overrightarrow{r}^{\prime }\right) }{\left| \overrightarrow{r}-%
\overrightarrow{r}^{\prime }\right| }d^{3}rd^{3}r^{\prime },  \nonumber
\label{deN1}
\end{eqnarray}

We are studying T-odd addition to the shift of level energy. To find wave
functions one can use the first order of perturbation theory due to the
weakness of interaction $V_w$ (see (10)). Transtion currents in the
first order over $V_w$ can be found with these wavefunctions: 
\begin{equation}
j_{NF}=j_{NF}^0+j_{NF}^T,  \label{jnf}
\end{equation}
where $j_{NF}^0$ is the transition current in the absence of weak T-odd
interactions (when $V_w=0$ ), $j_{NF}^T$ is the correction to transition
current caused by the presence of $V_w$ and calculated using wavefunctions (10).

Let us consider in details the part of $\delta {\varepsilon}_{N}^{(1)}$
caused by currents $\overrightarrow{j}_{NN^{\prime}}$:

\begin{equation}
\delta {\varepsilon}_{N}^{\overrightarrow{j}}= \frac {1} {8\pi } \int \frac{{%
\overrightarrow{j}}_{NN^{\prime}} ( {\overrightarrow{r}}) {\overrightarrow{j}%
}_{N^{\prime }N} \left( \overrightarrow{r}^{\prime }\right) } {\left| 
\overrightarrow{r}-\overrightarrow{r}^{\prime }\right| } d^{3}rd^{3}r^{%
\prime }  \label{deNj}
\end{equation}

Substitution of (\ref{jnf}) to (\ref{deNj}) gives

\begin{eqnarray}
\delta \varepsilon _{N}^{(\overrightarrow{j})}\approx \frac{1}{8\pi }\int 
\frac{{\overrightarrow{j}_{NN^{\prime }}^{0}}\left( \overrightarrow{r}%
\right) {\overrightarrow{j}_{N^{\prime }N}^{0}}\left( \overrightarrow{r}%
^{\prime }\right) }{\left| \overrightarrow{r}-\overrightarrow{r}^{\prime
}\right| }d^{3}rd^{3}r^{\prime }+  \nonumber \\
+\frac{1}{8\pi }\int \frac{{\overrightarrow{j}_{NN^{\prime }}^{0}}\left( 
\overrightarrow{r}\right) {\overrightarrow{j}_{N^{\prime }N}^{T}}\left( 
\overrightarrow{r}^{\prime }\right) +{\overrightarrow{j}_{NN^{\prime }}^{T}}%
\left( \overrightarrow{r}\right) {\overrightarrow{j}_{N^{\prime }N}^{0}}%
\left( \overrightarrow{r}^{\prime }\right) }{\left| \overrightarrow{r}-%
\overrightarrow{r}^{\prime }\right| }d^{3}rd^{3}r^{\prime }  \label{deNsum}
\end{eqnarray}
The first term in (\ref{deNsum}) describes conventional T-even contribution
to radiation correction, while the second term is the T-odd contribution to
the radiation shift of the level $\delta \varepsilon ^{\overrightarrow{j}~T}$%
.

Remembering that a vector-potential $\overrightarrow{A}(\overrightarrow{r})$
satisfies an equation \cite{15}: 
\begin{equation}
\Delta \overrightarrow{A}=-\frac{1}{c}(\overrightarrow{j}), \text{(here }
\Delta \text{ is the Laplace operator)}
\end{equation}
one can obtain: 
\begin{equation}
\overrightarrow{A}(\overrightarrow{r})=-\frac{1}{4 \pi c} \int \frac {%
\overrightarrow{j}(\overrightarrow{r}^{\prime }) } {\left| \overrightarrow{r}%
-\overrightarrow{r}^{\prime }\right|} d^{3}r^{\prime}.
\end{equation}

Therefore, the expression similar to (\ref{}) can be written:

\begin{equation}
\delta \varepsilon _{N}^{(\overrightarrow{j})T}=\frac{1}{2c}\int [{%
\overrightarrow{j}_{NN^{\prime }}^{0}}\left( \overrightarrow{r}\right) {%
\overrightarrow{A}_{N^{\prime }N}^{T}}\left( \overrightarrow{r}\right) +{%
\overrightarrow{A}_{NN^{\prime }}^{T}}\left( \overrightarrow{r}\right) {%
\overrightarrow{j}_{N^{\prime }N}^{0}}\left( \overrightarrow{r}\right)
]\,d^{3}r  \label{deNjT}
\end{equation}


\begin{thebibliography}{99}
\bibitem{1} S.K. Lamoreaux, LANL e-print arXive: nucl-ex/0109014v4 (2002).

\bibitem{2} M.V. Romalis, W.C. Griffith, and E.N. Fortson, LANL e-print
arXive: hep-ex/0012001v1 (2000).

\bibitem{3} B.C. Regan, Eugene D. Commins, Cristian J. Schmidt and David
DeMille, \textit{Phys.Rev.Lett} \textbf{88}, n.071805-1 (2002).

\bibitem{4} J.J. Hudson, B.E. Sauer, M.R. Tarbutt, and E.A. Hinds, LANL
e-print arXive: hep-ex/0202014v2 (2002), \textit{Phys.Rev.Lett} \textbf{89},
n.023003 (2002).

\bibitem{5} F.L. Shapiro, \textit{Sov. Phys. Usp.} \textbf{11} (1968) 345.

\bibitem{6} B.V. Vasil'ev and E.V. Kolycheva, \textit{Sov. Phys. JETP} 
\textbf{47} (1978) 243.

\bibitem{7} V.G. Baryshevsky LANL e-print arXive: hep-ph/9912270v2 (1999).

\bibitem{8} V.G. Baryshevsky LANL e-print arXive: hep-ph/9912438v3 (2000).

\bibitem{n9} V.G. Baryshevsky, \textit{Phys. Lett.} \textbf{A177} (1993) 38.

\bibitem{n10} V.G. Baryshevsky, \textit{J. High Energy Phys.} \textbf{04}
(1998) 018.

\bibitem{9} Khriplovich I.B. Parity Nonconservation in Atomic Phenomena.
1991 (London: Gordon and Breach).

\bibitem{12} V. G. Baryshevsky and D. N. Matsukevich, \textit{Phys.Rev.} 
\textbf{A66} (2002) 062110 (LANL e-print arXive: hep-ph/0002040 (2000)).

\bibitem{new_lanl} V. G. Baryshevsky LANL e-print arXive: hep-ph/0307291
(2003).

\bibitem{13} Baryshevsky V.G., Baryshevsky D.V. Journ. Of Phys. B: At. Mol.
Opt. Phys. 27 (1994) 4421.

\bibitem{15} A.I. Akhiezer and V.B. Berestetskii, Quantum Electrodynamics 
(Interscience Publishers, New York, London, Sydney, 1965).
\end{thebibliography}
\end{document}